\documentclass{iopart}

\usepackage[hypertex]{hyperref}
\usepackage{srctex}
\usepackage{graphicx}
\usepackage{iopams}
\usepackage{amssymb}
\newcommand{\RS}{\scriptscriptstyle{\rm RS}}
\newcommand{\RSB}{\scriptscriptstyle{\rm 1RSB}}

\begin{document}

\title{ Spin one $p$-spin glass: the Gardner transition.}

\author{T.I. Schelkacheva}
\address{Institute for High Pressure Physics, Russian Academy of Sciences, Troitsk 142190,
Moscow Region, Russia}
\author{E.E. Tareyeva}
\address{Institute for High Pressure Physics, Russian Academy of Sciences, Troitsk 142190,
Moscow Region, Russia}

\begin{abstract}
We examine the phase diagram of the $p$-spin mean field glass model in the
spin one case, that is when $S=0,+1,-1$. For large $p$ the model is solved
exactly. The analysis reveals that the phase diagram is in some way
similar to that of Ising spins. However, as we show, the quadrupolar
regular ordering as well as qudrupolar glass order are present now.
The temperature of the Gardner 1RSB -- FRSB transition is obtained
explicitly for large $p$. The case of higher spins is discussed briefly.
\end{abstract}

\maketitle

\section{Introduction\label{Sec:Intro}}
The $p$-spin spin-glass model of $p$ randomly interacting Ising spins was
introduced \cite{GrosMezard,Gardner} as a natural generalization of the
Sherrington-Kirkpatrick model \cite{sk}. In the  $p$-spin model there
exists temperature interval where the first step replica symmetry breaking
(1RSB) solution is stable. In mean field pure $p$-spin spherical glasses \cite{Thirumalai}
this interval extends to zero temperature and the transition from replica
symmetric (RS) to 1RSB solution is jumpwise in the glass order parameter.
It is not so the case when the model contains terms with different
values of $p$ \cite{Leuzzi}. The papers on 1RSB spin glasses make now the basis
for so called equilibrium approaches to real glasses (see for reviews
\cite{ParZamp,Wol,biroli}).

In discontinuous mean field Ising spin $p$-spin glass of \cite{GrosMezard,Gardner}
the mentioned interval of 1RSB stability is finite and increases with
increasing $p$. It is zero for $p=2$ \cite{sk}. The low-temperature
boundary of the 1RSB stability region is given by the so called Gardner
transition temperature \cite{Gardner}. Recently it was shown
\cite{Par1,Par2,UrBi} that the Gardner transition plays an important role
in the equilibrium theory of structural glasses at high pressures near
jamming transition. Jamming transition takes place at pressures higher
than the Gardner pressure. One can say that at high pressure it is just
non-spherical $p$-spin glasses that are the prototypes of glassy behavior.
At very low temperatures in the case of spin glasses or at very high
pressure in the case of structural glasses the slow dynamics is determined
by the state landscape of FRSB \cite{MoRT}.

The detailed work by Rizzo \cite{Rizzo} and our own experience
in the investigation of arbitrary operator p-operator spin-glass-like
models \cite{SchelkT,Sc,Schelkacheva,2014a,2014b} as well as of Potts
spin-glass models \cite{Potts1,Potts2} bring us to the conclusion that
Gardner transition is the phenomenon which is common for a large class of
models, although explicit analytical result, as far as we know, was
obtained in only one case \cite{Gardner}. So, in the present paper we
investigate this problem in the case of the spin one $p$-spin glass.

The $p$-spin spin glass with $S=0,\pm 1$ presents the particular case of
the problems considered in papers \cite{M,A} and so a part of our results
we obtain now repeat theirs. However, the authors of \cite{M,A} have paid
no attention to the stability of 1RSB solution and the Gardner transition
and to the existence of quadrupolar glass in the case of pure $p$-spin
interaction. The investigation of these two problems presents a goal of
the present paper.

The paper is organized as follows. In sect.2.1 the model is described and
the main equations obtained by replica approach are obtained. In sect.2.2
the case of infinite and that of large $p$ are investigated. Replica
symmetric solution as well as the first step replica symmetry breaking
solution are considered. The stability of 1RSB solution is examined for
large $p$. The low-temperature instability (Gardner transition) point is
obtained explicitly in analytical way. In sect.2.3 for large $p$ the
existence of quadrupole orientational glass is demonstrated. In sect.3 the
analogous results for higher spin models are derived and discussed.

\section{Spin one p-spin glass model \label{Sec:glass}}

\subsection{Main equations}

Let us consider the $p$-spin model with the Hamiltonian
\begin{equation}
H=-\sum_{{i_{1}}\leq{i_{2}}...\leq{i_{p}}}J_{i_{1}...i_{p}}
\hat{S}_{i_{1}}\hat{S}_{i_{2}}...\hat{S}_{i_{p}}, \label{one}
\end{equation}
Here $\hat{S}$ now is the diagonal  spin one operator ($S=0, +1, -1 $),
$N$ is the number of lattice sites, $i=1,2,...N$, and $p$ is the
number of interacting particles.
$J_{i_{1}...i_{p}}$ are independent random variables with  Gaussian
distribution
\begin{equation}
P(J_{i_{1}...i_{p}})=\frac{\sqrt{N^{p-1}}}{\sqrt{p!\pi}
J}\exp\left[-\frac{(J_{i_{1}...i_{p}})^{2}N^{p-1}}{ p!J^{2}}\right]. \label{two}
\end{equation}

Using replica approach we can write the free energy averaged over disorder
in the following form that we write here because it is  instructive to
compare our case with the free energy of the random $p$-spin model in the
case of Ising spins \cite{GrosMezard,Gardner}:

\begin{eqnarray}
\nonumber
\frac{F}{NkT} = \lim_{n \rightarrow 0\alpha } \frac{1}{n} max
\left[-\frac{t^2}{4} \sum _{\alpha }\left(\frac{2+y_{\alpha
}}{3}\right)^{p} +t^{2}\sum _{\alpha } \mu _{\alpha
}\frac{2+y_{\alpha  }}{3} - \right.
\\
\left.\frac{t^{2}}{4}\sum _{\alpha \neq \beta } (q^{\alpha
\beta })^{p}+ t^{2}\sum _{(\alpha \beta )} q^{\alpha \beta }\lambda
_{\alpha
\beta}  - \ln \Tr \exp \hat{\theta}\right ]
\label{fab}
\end{eqnarray}

with
$$\hat {\theta} = t^{2} \sum _{(\alpha \beta )} \lambda _{\alpha
\beta } \hat {S}_{\alpha } \hat {S}_{\beta } + t^{2}\sum _{\alpha
} \mu _{\alpha}  \frac{2+\hat {Q}_{\alpha }}{3}.$$
$\hat {Q}$ is the quadrupole-moment operator,
$$ (\hat{S}_{\alpha })^{2} = \frac{2+ \hat {Q}_{\alpha }}{3}$$
in the case $S =1$, $y_{\alpha } = <\hat{Q}_{\alpha }>$ is the regular
quadrupole order parameter and $q^{\alpha \beta }$ is the spin glass order
parameter; $\lambda,\mu$ are Lagrange multipliers, 
$t={J}/kT$.

Below we will use another form which follows explicitly from our
 papers where the detailed
calculations for the case of interaction of $p$ arbitrary diagonal
operators $\hat{U}$ are given  (see, e.g.\cite{SchelkT}). Let us  perform
the first stage of the replica symmetry breaking (1RSB)  ($n$ replicas are
divided into  $n/m_{1}$ groups with  $m_{1}$ replicas in each group) and
obtain the expression for the free energy. Glass order parameters are
denoted by $q^{\alpha
\beta }= q_{0}$ if $\alpha $ and
$\beta $ are from different groups and $q^{\alpha \beta }= q_{1}$ if
$\alpha $ and $\beta $ belong to the same group. So
\begin{eqnarray}
\nonumber\fl F_{\rm \RSB}=-NkT\left\{m_{1} t^2(p-1)\frac{q_{0}^p}{4}+(1-m_{1})(p-1)
t^2\frac{(q_{1})^p}{4}-t^2(p-1)\frac{{w_{1}}^p}{4}+ \right.
\\
\left. \frac{1}{m_{1}}\int dz^G\ln
\int ds^G
\left[\Tr\exp\left(\hat{\theta}_{\RSB}\right)\right]^{m_{1}}\right\}.
\label{frs}
\end{eqnarray}

Here
\begin{eqnarray}\label{theta}
\nonumber\fl \hat{\theta}_{\rm \RSB}=\left.zt\sqrt{\frac{p{q_{0}}^{(p-1)}}{2}}\,\hat{S}+st\sqrt{\frac{p[{(q_{1})}^{(p-1)}-{q_{0}}^{(p-1)}]}{2}}\,\hat{S}+
\right.
\\
\left.
t^2\frac{p[{w_{1}}^{(p-1)}-{(q_{1})}^{(p-1)}]}{4}\hat{S}^2,\right.
\end{eqnarray}
\begin{equation}
\int dz^G = \int_{-\infty}^{\infty} \frac{dz}{\sqrt{2\pi}}\exp\left(-\frac{z^2}{2}\right).
\end{equation}
The extremum conditions for  $F_{\rm \RSB}$ yield equations for the order
parameters. We get glass order parameters $q_{0}$ and $q_{1}$, the
auxiliary order parameter $w_{1}$, the regular order parameter $x_{1}$ and
the parameter $m_{1}$. Auxiliary order parameter $w_{1}$ arises from the
fact that $\hat{S}$ in Eq.~(\ref{one}) are not Ising spins.

\begin{eqnarray}\label{18qrs}
q_{0}=\int dz^G\left\{ \frac{\int ds^G{\left[\Tr\exp\hat{\theta}_{\rm \RSB}\right]}^{(m_{1}-1)}\left[\Tr\hat{S} \exp\hat{\theta}_{\rm \RSB}\right]}
{\int ds^G{\left[\Tr\exp\hat{\theta}_{\rm \RSB}\right]}^{m_{1}}}\right\}^{2},
\\\nonumber\fl
q_{1}=
\int dz^G \frac{\int ds^G{\left[\Tr\exp\hat{\theta}_{\rm \RSB}\right]}^{(m_{1}-2)}{\left[\Tr{\hat{S} }\exp\hat{\theta}_{\rm \RSB}\right]}^{2}}
{\int ds^G{\left[\Tr\exp\hat{\theta}_{\rm \RSB}\right]}^{m_{1}}},
\\\label{1prs}
w_{1}=
\int dz^G \frac{\int ds^G{\left[\Tr\exp\hat{\theta}_{\rm \RSB}\right]}^{(m_{1}-1)}\left[\Tr{\hat{S} }^{2}\exp\hat{\theta}_{\rm \RSB}\right]}
{\int ds^G{\left[\Tr\exp\hat{\theta}_{\rm \RSB}\right]}^{m_{1}}},
\\
x_{1}=
\int dz^G \frac{\int ds^G{\left[\Tr\exp\hat{\theta}_{\rm \RSB}\right]}^{(m_{1}-1)}\left[\Tr{\hat{S} }\exp\hat{\theta}_{\rm \RSB}\right]}
{\int ds^G{\left[\Tr\exp\hat{\theta}_{\rm \RSB}\right]}^{m_{1}}}.
\label{4prs}
\end{eqnarray}
Similarly, we obtain the equation for the order parameter $m_{1}$.

The corresponding expressions  for the RS approximation can be easily
obtained from the preceding formulas (\ref{frs})-(\ref{4prs}) when
$q_{0}=q_{1}$. We have
\begin{equation}\label{0qrs}
q_{\rm \RS}=\int dz^G\left\{ \frac{\Tr\left[\hat{S} \exp\left(\hat{\theta}_{\rm \RS}\right)\right]}
{\Tr\left[\exp\left(\hat{\theta}_{\rm \RS}\right)\right]}\right\}^{2},
\end{equation}
\begin{equation}\label{2qrs}
w_{\rm \RS}=\int dz^G\frac{\Tr\left[{\hat{S}}^{2} \exp\left(\hat{\theta}_{\rm \RS}\right)\right]}
{\Tr\left[\exp\left(\hat{\theta}_{\rm \RS}\right)\right]},
\end{equation}
\begin{equation}\label{3qrs}
x_{\rm \RS}=\int dz^G\frac{\Tr\left[{\hat{S}}\exp\left(\hat{\theta}_{\rm \RS}\right)\right]}
{\Tr\left[\exp\left(\hat{\theta}_{\rm \RS}\right)\right]},
\end{equation}
Here
\begin{equation}\label{1qrs}
\hat{\theta}_{\rm \RS}=zt\sqrt{\frac{p\,{q_{\rm \RS}}^{(p-1)}}{2}}\,\hat{S}+t^2\frac{p[{w_{\rm \RS}}^{(p-1)}-
{q_{\rm \RS}}^{(p-1)}]}{4}\hat{S}^2.
\end{equation}

\subsection{Large $p$ solutions. Gardner transition temperature.}

In the case of $p\rightarrow  \infty$  the problem can be solved exactly
~\cite{M,A}. Consideration of such a limiting case makes it possible to
describe many properties of the model for finite values of $p$.

 It is easy to see that order parameters come in  $\hat{\theta}$ and $F$ in
the form of a power function $q^{p}$ and $w^{p}$. Herewith $0\leq q\leq1$
and $0\leq w\leq1$.

Let us consider first the replica symmetric case in the limit
$p\rightarrow  \infty$. Let's pretend that $0\leq q<1$ and $0\leq w<1$.
Then $q^{p}=w^{p}=0$ and we get directly $q_{\rm \RS}=0$, $x_{\rm \RS}=
<\hat{S}> = 0$ and $w_{\rm \RS}=2/3$ from Eq.~(\ref{0qrs}) -
Eq.~(\ref{1qrs}). The ordering of spins is absent.

Quadrupole operator $\hat{Q}=3{\hat{S}}^{2}-2=(-2,1,1)$. So a regular
quadrupolar ordering is also absent because average value $<\hat{Q}> =
3w_{\rm \RS}-2 = 0 $. We have got disordered paraphase with the free
energy $F/(NkT)=-\ln 3$.

There is another solution in the replica  symmetric consideration. When
$0\leq q<1$ and $w=1$ we have $q^{p}=0, w^{p}=1 $. It turns out from
Eq.~(\ref{0qrs}) - Eq.~(\ref{1qrs}) that $q_{\rm \RS}=0$, $x_{\rm \RS}=
0$ and $w_{\rm \RS}=1$. Then average values $<\hat{Q}> = 3w_{\rm \RS}-2 =
1 $ and $<\hat{S}> = 0$. The phase is not ordered in spins, but there is a
quadrupole ordering.

It is important to note that the contribution  to ordering is
given only by the states $S=+1,-1$. The free energy is:
\begin{equation}
\fl F_{\RS}/(NkT)=- J^{2}/(2kT)^{2}-\ln 2.\label{a}
\end{equation}
It is identical  to the result for the case of Ising spin $\hat{S}=(+1,-1)$
~\cite{GrosMezard,Gardner}.

All these RS states as well as the transitions between them are
described in details in \cite{M,A}.

Let us consider now the 1RSB case in the limit $p\rightarrow  \infty$.
Let us emphasize that now it is the paraphase with the nonzero quadrupolar
 ordering that bifurcates.In accordance with the Parisi approach ~\cite{P}
 we have $q_{0}<q_{1}$. Hence we  immediately obtain the order parameters
 from Eq.~(\ref{18qrs}) - Eq.~(\ref{4prs}): $q_{0}=0$, $q_{1}=1$,
$x_{1}= <\hat{S}> = 0$ and $w_{1}=1$. So we have got a glass phase
with a nonzero spin glass order parameter $q_{1}=1$. The quadrupole
ordering is preserved $<\hat{Q}> = 3w_{1}-2=1$. Still contribution is only
from $S=+1,-1$. A value of $S=0$ does not make a contribution.

 The expression for the free energy is given by (Eq.~(\ref{frs})) and has
the form:
\begin{equation}
\fl F_{\rm \RSB}/(NkT)=- m_{1}J^{2}/(2kT)^{2}-(1/m_{1})\ln 2.\label{B}
\end{equation}
 It coincides  with that which was obtained for the Ising spins
~\cite{GrosMezard,Gardner}. The expression for $m_{1}$ can be obtained as
the extremum condition for $F_{\rm \RSB}/(NkT)$:
\begin{equation}
{m_{1}}^{2}J^{2}/(2kT)^{2}=\ln 2.\label{m}
\end{equation}
When $m_{1}=1$  free energies Eq.~(\ref{a}) and Eq.~(\ref{B}) become equal.
From Eq.~(\ref{m}) we have $kT_{c}/J=1/(2\sqrt{\ln 2})$.   Since
${m_{1}}J/(2kT)$ is independent of temperature. $F_{\rm \RSB}$ is
independent of temperature too below $T_{c}$. This is exactly the same
results as for the problem with Ising spins ~\cite{GrosMezard,Gardner}.
This 1RSB solution was obtained in \cite{A}. The problem of its stability
was not considered in that paper. Let us investigate it now.

We can break the replica symmetry  in our model Eq.~(\ref{one}) once  more
and obtain the corresponding expressions for the free energy and the order
parameters. The bifurcation condition  $\lambda_{\rm (\RSB) repl}=0$
determining the temperature $T = T_{G}$ (the Gardner temperature) of
instability follows from the condition that a nontrivial small solution
for the 2RSB glass order parameter appears (see ~\cite{Sc}). We have:

\begin{eqnarray}
\nonumber\fl\lambda_{\rm (\RSB) repl}= 1 - t^{2}
 \frac{p(p-1)(q_{1})^{(p-2)}}{2} \times \\
\fl\int dz^G \frac{\int ds^G\left[\Tr\exp\left(\hat{\theta}_{\RSB}\right)\right]^{m_{1}}
\left\{\frac{\Tr\left[\hat{S}^2
\exp\left(\hat{\theta}_{\RSB}\right)\right]} {\Tr\left[\exp\left(\hat{\theta}_{\RSB}\right)\right]}-
\left[\frac{\Tr\left[\hat{S} \exp\left(\hat{\theta}_{\RSB}\right)\right]}
{\Tr\left[\exp\left(\hat{\theta}_{\RSB}\right)\right]}\right]^2\right\}^2.}{\int ds^G
\left[\Tr\exp\left(\hat{\theta}_{\RSB}\right)\right]^{m_{1}}}\label{lambda}
\end{eqnarray}
Eq.~(\ref{lambda})  depends only on  1RSB-solution. It has been shown that
1RSB solution is stable when $\lambda_{\rm (\RSB) repl}>0$
~\cite{Gardner,SchelkT}.

When $p\rightarrow \infty$ the  1RSB glass solution is stable, because we
have $(2kT)^{2}\lambda_{\rm (\RSB) repl}>0$ from Eq.~(\ref{lambda}) at all
temperatures $T>0$.
At large but finite values of  $p$ the condition $\lambda_{\rm (\RSB)
repl}>0$ is violated at low temperature $T_{G} \neq 0$.

Let us calculate the Gardner transition temperature $T_{G}$ explicitly
from the requirement  $\lambda_{\rm (\RSB) repl}=0$. In our case
$\hat{\theta}_{1RSB} = st \sqrt{\frac{p}{2}}$ and
\begin{equation}
\Psi(s) \equiv \Tr \exp \hat{\theta}_{1RSB} = 1+ 2 \cosh
st\sqrt{\frac{p}{2}}.
\label{psi}
\end{equation}

We can rewrite the equation for $\lambda $ in the form:

\begin{equation}
\lambda = 1 - \frac{t^{2}}{2} p (p-1) \frac{\int
ds^{G}\Psi(s)^{m-4}(3+\Psi(s))^{2}}{\int ds^{G}\Psi(s)^{m}}
\label{lamnow}
\end{equation}
Calculating the transition point we keep in mind that at large $p$ the
values of $T$ and $m$ are small. So we obtain after deriving the
asymptotics of integrals and summing the series the equation for the
limit of stability of 1RSB phase:
\begin{equation}
1 = \frac{p^{3/2}t}{4 \sqrt{\pi} 2^{p}} \left(\frac{5}{6}-\frac{2 \pi}{9
\sqrt{3}}\right),
\nonumber
\end{equation}
so, that
\begin{equation}
kT_{G}/J=\frac{0.1076}{\sqrt{\pi}} \frac{p^{3/2}}{2^{p}}.
\label{tgard}
\end{equation}
Let us note that the $p$-dependence is of the same form as in the case of
Ising spins.

\subsection{Large $p$ solution. The existence of the quadrupole glass.}

 The main difference of Ising and spin one cases is the  presence of a
quadrupole ordering in the latter one. We can define  the quadrupole
glass (orientational) order parameter by
Eq.~(\ref{18qrs}) and Eq.~(\ref{1prs}) replacing $\hat{S}$ to $\hat{Q}$
and keeping in mind the zero limit of interaction of quadrupoles . The
function $\hat{\theta}_{\rm \RSB}$ is not changed at the shutdown of
quadrupole-quadrupole interaction.
\begin{eqnarray}\label{38qrs}
q_{0}^{Q}=\int dz^G\left\{ \frac{\int ds^G{\left[\Tr\exp\hat{\theta}_{\rm \RSB}\right]}^{(m_{1}-1)}\left[\Tr\hat{Q} \exp\hat{\theta}_{\rm \RSB}\right]}
{\int ds^G{\left[\Tr\exp\hat{\theta}_{\rm \RSB}\right]}^{m_{1}}}\right\}^{2},
\\\nonumber\fl
q_{1}^{Q}=
\int dz^G \frac{\int ds^G{\left[\Tr\exp\hat{\theta}_{\rm \RSB}\right]}^{(m_{1}-2)}{\left[\Tr{\hat{Q} }\exp\hat{\theta}_{\rm \RSB}\right]}^{2}}
{\int ds^G{\left[\Tr\exp\hat{\theta}_{\rm \RSB}\right]}^{m_{1}}}.
\\\label{3prs}
\end{eqnarray}

   In the limit of infinite $p$ we obtain quadrupole  glass order
parameters ${q_{0}}^{Q}=1$ and ${q_{1}}^{Q}=1$. But we can not say that,
along with the spin glass we obtain a quadrupole glass, because
  $(<\hat{Q}>_{1RSB})^{2}={q_{0}}^{Q}={q_{1}}^{Q}=1$.

To clarify the question of the presence or absence of quadrupole glass
let us consider now the case of large but finite values of $p$.
It is suitable  to write the equations for $<\hat{Q}>_{1RSB}$ and $q_{1}^{Q}$ as
follows:

$$<\hat{Q}>_{1RSB}= 1 - 3 \frac{\int ds^G \Psi(s)^{m-1}} {\int ds^G
\Psi(s)^{m}}
$$

$$q_1^Q = 1 - 6 \frac {\int ds^G \Psi(s)^{m-1}} {\int ds^G \Psi(s)^{m}}  +
9 \frac {\int ds^G \Psi(s)^{m-2}} {\int ds^G \Psi(s)^{m}}  $$
and we have

\begin{equation}
q_{1}^{Q} - <\hat{Q}>_{1RSB}^{2} = 9\left[
 \frac {\int ds^G \Psi(s)^{m-2}} {\int ds^G \Psi(s)^{m}}  -
 \left(\frac {\int ds^G \Psi(s)^{m-1}} {\int ds^G
\Psi(s)^{m}}\right)^{2}\right].
\label{delta}
\end{equation}
Now we can proceed as when obtaining the Gardner temperature. At large
$p$ the integrals $\int ds^G \Psi(s)^{\eta }$ with $\eta >0$ are
approximately equal to $2*2^{p}$ while those for $\eta < 0$ are
proportional to $1/\sqrt{\pi p}$. So, the considered difference is
\begin{equation}
q_{1}^{{Q}} - <\hat{Q}>_{1RSB}^{2} = 9\left[\frac{\Sigma_{1}}{\sqrt{\pi p} 2^{p}} -
\left(\frac{\Sigma_{2}}{\sqrt{\pi p} 2^{p}}\right)^{2}\right] >0
\label{delta1}
\end{equation}
with $\Sigma_{i}$ standing for converging sums that can be easily
evaluated.

This means that for large but finite $p$ the
quadrupolar orientational glass is present along with the spin glass in
spin one system.

Such a phenomenon was first encountered in generalized  SK ~\cite{sk} model
for spin one case investigated in RS approximation in the paper
 Ref.~\cite{LT}. After performing high-temperature series expansion of the
 RS equations Eq.~(\ref{0qrs}) - Eq.~(\ref{1qrs}), one easily makes sure
 that average value of quadrupole $<\hat{Q}>_{RS} = 3w_{\rm \RS}-2$ is
different from zero at arbitrarily high temperatures, so random
distribution of spins $\hat{S}$ produces a non-zero average value of the
quadrupole moment, which gradually increases with decreasing
temperature to a critical temperature $T_{c}$. At temperatures below
$T_{c}$, the system continues to have quadrupole ordering. At the
point $T_{c}$ appears spin glass and quadrupole glass, too, i. e., the order
parameter defined in the replica symmetric consideration by the relation
\begin{equation}\label{30qrs}
q_{\rm \RS}^{Q}=\int dz^G\left\{ \frac{\Tr\left[\hat{Q} \exp\left(\hat{\theta}_{\rm \RS}\right)\right]}
{\Tr\left[\exp\left(\hat{\theta}_{\rm \RS}\right)\right]}\right\}^{2},
\end{equation}
ceases to be equal to $[<\hat{Q}>_{RS}]^{2}$. We define the parameters of the quadrupole glass
in the limit off quadrupole interaction.

Low-temperature asymptotic behavior of the order parameters may be obtained
analytically:
\begin{equation}\label{40qrs}
<\hat{Q}>_{RS}=1+O(e^{-t}), q_{\rm \RS}^{S}=1-\frac{2}{3\sqrt{2}\pi}t^{-1}+O(e^{-t}).
\end{equation}

The most interesting of these results is the fact
that $<\hat{Q}>_{RS}=1$ and $q_{\rm \RS}^{S}=1$ at $T = 0$. This means that at zero temperature, all
spins take only the values +1 and -1. The state $S=0$ is absent. It may be used to determine
the number of metastable states $<N_{S}>=e^{0.19923N}$ at zero temperature. It is known result for the SK model ~\cite{T}.
So that our spin one glass is quite similar to the Ising spin
glass at $T=0$.

\section{Higher spins $p$-spin glass \label{Sec:High}}

Consider now the case of Hamiltonian Eq.~(\ref{one})  with larger spin
values $j=2,3,...$.  We will use normalized operators
$\hat{S}={\hat{j}}_{z}/j$ that is much more convenient for calculations and
does not change the symmetry of the problem. So when $j=1$ we have
$(1,0,-1)$ as before. For $j=2$ we use
$\frac{1}{2}{\hat{j}}_{z}=\frac{1}{2}(2,1,0,-1,-2)$. For $j=3$ we use
$\frac{1}{3}{\hat{j}}_{z}=\frac{1}{3}(3,2,1,0,-1,-2,-3)$ and so on.

As is known, the quadrupole moment is
${\hat{Q}}\sim[3{\hat{j}}_{z}^{2}-j(j+1)]$ in the space $j=const$. We will
use normalized expression for quadrupole moment
$[3{\hat{j}}_{z}^{2}-j(j+1)]/j^{2}$ for uniformity of computing
that does not change the symmetry of the problem.

We operate on completely similar to  the previous case of spin one. First
of all, we get completely disordered paraphase. We have not glass
$q_{RS}^{S}=0 $. There is no ordering of the spins. We obtain $
0<w_{RS}<1$: $w_{RS}=2/3$ for $j=1$, $w_{RS}=1/2$ for $j=2$, $w_{RS}=4/9$
for $j=3$, $w_{RS}=5/12$ for $j=4$. For average valule
$<{\hat{Q}}_{j}>=3<({\hat{j}}_{z}/j)^{2}>- (j+1)/{j} =3w_{RS}-(j+1)/{j}$
we get $ <{\hat{Q}}_{j=1}>=<{\hat{Q}}_{j=2}>=<{\hat{Q}}_{j=3}>=...=0 $.
Hence, we have no quadrupole ordering. Free energy is $F/(NkT)=-\ln
(2j+1)$, so $F/(NkT)=-\ln (3)$ for $j=1$, $F/(NkT)=-\ln (5)$ for $j=2$,
and so on. We obtain from $q_{RS}<1$ and  Eq.~(\ref{1qrs}) that
$\hat{\theta}_{\rm \RS}=t^2\frac{p{w_{\rm \RS}}^{(p-1)}}{4}\hat{S}^2$
where $\hat{S}=\hat{j_{z}} /{j}$. So we get from  $w_{RS}<1$ and
Eq.~(\ref{2qrs}) that completely disordered phase  occurs at a purely
formal mathematical case $p\rightarrow\infty$. For arbitrarily large but
finite values of $p$ such a phase takes place only at $T\rightarrow\infty$
(or $t\rightarrow 0$).

 The results presented below present a strict generalization of that for
spin one case. The formulas  Eq.~(\ref{a}) -
Eq.~(\ref{m}) hold now, too. This is due to the fact that in calculating
the integrals of the type $\int ds^{G} [\Tr
\exp(s\sqrt{\frac{p}{2}}t\hat{S})]^{m}$ only the terms with the
largest absolute values of the normalized operator
$\hat{S}=\hat{j_{z}/}{j}$ do contribute. We obtain two
low-temperature phases: disordered phase of spin values ($q_{\rm \RS}=0$,
$x_{\rm \RS}=<\hat{j_{z}/}{j}>=  0$, $w_{\rm \RS}=1$) and 1RSB spin glass
phase ($q_{0}^{s}=0$, $q_{1}^{s}=1$, $x_{1}^{s}= <\hat{j_{z}/}{j}>_{\RSB}=
0$ and $w_{1}=1$). The first of these phases (no glass) for $j\geq 2$ is
formally less energetically favorable than paraphase when the temperature
decreases from arbitrarily high temperature to a temperature of occurrence
of 1RSB spin glass.

 Quadrupole ordering occurs  in two low-temperature phases: average values
are
$<{\hat{Q}}_{j=1}>=1$, $<{\hat{Q}}_{j=2}>=3/2$,  $ <{\hat{Q}}_{j=3}>=5/2$
. In 1RSB spin glass phase we obtain from Eq.~(\ref{38qrs}) -
Eq.~(\ref{3prs}) quadrupole glass order parameters
${q_{0}}^{Q}={q_{1}}^{Q}=[<{\hat{Q}}_{j=1}>]^{2}=1$ for $j=1$,
${q_{0}}^{Q}={q_{1}}^{Q}=[<{\hat{Q}}_{j=2}>]^{2}=[3/2]^{2}$ for $j=2$,..
This is a consequence of the fact that only the maximum values of the
operator $\hat{Q}$ significantly contribute to $\Tr
\hat{Q}exp(\hat{\theta}_{\rm \RSB})$ under the integral  when
$p\rightarrow \infty$. So there is no quadrupole glass along with the spin
glass in the limit $p\rightarrow \infty$ limit.


\section{Conclusions \label{Sec:Conc}}

The phase diagram of the $p$-spin mean field glass model in the
spin one case is examined in details for large values of $p$. Some new
facts, as compared with \cite{M,A}, are established. It is shown
that 1RSB phase is unstable at low temperatures and the low-temperature boundary of stability,
the Gardner temperature, is calculated explicitly in analytical way.
An interesting feature of spin glass phase, as we show, is the
existence
 of quadrupole orientational glass along with the spin glass in
addition to known regular quadrupole ordering.
The case of higher spins is discussed, too.

It is interesting that the contribution to
the values of order parameters comes from all the values of $\hat{S}$ only
at high temperatures. At low temperatures, the main contribution is made
by the maximum values of the spin operators. It is possible that for other
operators and models of cluster glass \cite{crst,tmf} some components or some of the
spherical harmonics effectively fall out. Perhaps the
successful description of the various complex glass systems through a
simple Ising spins can be explained by this fact.
Note that a complication of interaction  (compared to Ising
spins) leads to induced hidden regular order and hidden glass
for the operators that do not enter the Hamiltonian.

\section{Acknowledgments}

This work was supported in part by the Russian Foundation
for Basic Research ( No. 14-02-00451).

\end{document}